\documentclass[prd,showpacs]{revtex4}
\begin{document}

\newcommand{\be}{\begin{equation}}
\newcommand{\ee}{\end{equation}}
\newcommand{\bea}{\begin{eqnarray}}
\newcommand{\eea}{\end{eqnarray}}


\title{Null string evolution in black hole and cosmological spacetimes}

\author{Mariusz P. D\c{a}browski}
\email{mpdabfz@uoo.univ.szczecin.pl}
\affiliation{\it Institute of Physics, University of Szczecin, Wielkopolska 15,
          70-451 Szczecin, Poland}
\author{Izabela Pr\'ochnicka}
\email{ipro@ift.uni.wroc.pl}
\affiliation{\it Institute of Theoretical Physics,
University of Wroc{\l }aw, pl. Maxa Borna 1, 50-205 Wroc{\l }aw, Poland}

\date{\today}

\input epsf

\begin{abstract}
We discuss the problem of the motion of classical strings in some black hole and cosmological
spacetimes. In particular, the null string limit (zero tension) of tensile strings is considered.
We present some new exact string solutions in Reissner-Nordstr\"om black hole
background as well as in the Einstein Static Universe and in the Einstein-Schwarzschild
(a black hole in the Einstein Static Universe) spacetime. These solutions can give some insight
into a general nature of propagation of strings (cosmic and fundamental) in curved backgrounds.
\end{abstract}

\pacs{11.25.Mj, 04.70.Bw, 04.62.+v, 98.80.Hw}
\maketitle

\section{Introduction}
\label{intro}

Fundamental string theory is undoubtedly the most serious
candidate for unification of gauge interactions with gravity
\cite{green}. Its effects should clearly be visible in extremely
high gravitational fields of black holes and in the early
universe. It is not an easy task to study quantum string
propagation in these background fields and this gives motivation
to study the motion of classical strings in these fields first in
order to catch uo some really ``stringy'' properties of a quantum
theory. On the other hand, classical motion of strings gives an
appropriate formalism to study the dynamics of cosmic strings
which appear naturally in GUT models \cite{vilenkinbook}. This is
why we will study the classical motion of strings in some black hole
and cosmological spacetimes.

Classical motion of strings which evolve in curved spacetimes can be described by a
system of the second-order non-linear coupled partial differential
equations \cite{vesan,integr}. The non-linearity of these equations gives a complication
which leads to their non-integrability and possibly chaos \cite{ott}.
It is well-known that various types of nonlinearities appear in
Newtonian as well as relativistic systems and so they can deliver chaos.
On the other hand, some types of non-linear equations can be
integrable and their solutions are not chaotic. It
seems that theory of relativity is ideal to produce chaotic
behaviour since their basic equations are highly non-linear.
However, the problem is not as easy as one could think of, because
most of the systems under study possess some symmetries which
simplify the problem. This also refers to a single particle
obeying either Newtonian or relativistic equations. Simply, a
single particle which moves in the gravitational field of a source
of gravity cannot move chaotically. However, two particles which form a
3-body system including the source can move in a chaotic way,
though still not for all possible configurations.

Admission of extended objects such as strings gives another
complication which, roughly, can be compared to the fact that now
we have a many-body system which can obviously be chaotic on the
classical level. An extended character of a string is reflected by
the equations of motion which become a very complicated non-linear
system from the very beginning. Thus, no wonder chaos can appear
for classical evolution of strings around the simplest sources of
gravity such as Schwarzschild black holes. This, in fact, was
explicitly proven \cite{lar94,frolar}. However, in a similar way
as for other types of non-linear sets of equations, there exist
integrable configurations. The investigation of such explicit
configurations can give an interesting insight into the problem of
the general evolution of extended objects in various sources of
gravity. Of course, it is justified, provided we do not consider
back-reaction of these extended objects onto the source field,
i.e., if we consider test strings in analogy to test particles
which do not ``disturb'' sources' gravitational fields.

Studies of exact configurations can give big insight into the
problem. One useful example is when unstable periodic orbits (UPO)
appear. Their emergence becomes a signal for a possible chaotic
behaviour of the general system \cite{frankel}.

The task of this paper is to study some exact configurations for
strings moving in simple spacetimes of general relativity.
Unfortunately, for strings, the main complication
refers to their self-interaction reflected in the
equations of motion by a non-zero value of tension (tensile
strings). However, one is able to study simpler extended
configurations for which tension vanishes called null
(tensionless) strings \cite{null,venico,LS96}. Their
equations of motion are null geodesic equations of general relativity
appended by an additional `stringy' constraint. Many exact null string
configurations in various curved spacetimes have already been studied
\cite{LS96,kar,dablar,mar2,alex,porfyriadis98,kuirokidis99,jassal}. One of the advantages of the null
string approach is the fact that one may consider null strings as
null approximation in various perturbative schemes for tensile
strings \cite{venico,alex1,alex2,alex3}.

In section II we present tensile and tensionless string equations
of motion. In Section III we obtain exact null string configurations
both in Reissner-Nordstr\"om and Schwarzschild spacetime while in
Section IV we derive string configurations in static Einstein
Universe. In Section V we discuss the evolution of strings in
Einstein-Schwarzschild (Vadiya) Universe. In Section VI we discuss our solutions.

\section{Tensile and null strings in curved spacetimes}
\label{sectionII}

A free string which propagates in a flat Minkowski spacetime sweeps
out a world--sheet (2-dimensional surface) in contrast to a point
particle, whose history is a world-line. The world--sheet action for
a free, closed string is given by the formula \cite{polch}
\begin{equation}
\label{sheet}
S= \frac{T}{2} \int d\tau d\sigma  \sqrt{-h}
h^{ab} \eta_{\mu\nu} \partial_{a} X^{\mu} \partial_{b}
X^{\nu}   ,
\end{equation}
where $T=1/2\pi\alpha'$ is the string tension, $\alpha'$ the Regge slope,
$\tau$ and $\sigma$
are the (spacelike and timelike, respectively) string coordinates,
$h^{ab}$ is a 2-dimensional world--sheet metric $(a,b = 0,1)$, $h = \det(h_{ab})$,
$X^{\mu}(\tau,\sigma)$
($\mu, \nu = 0, 1, \ldots D - 1)$ are the coordinates of the string world--sheet
in D-dimensional Minkowski spacetime with metric
$\eta_{\mu\nu}$.

If instead of the flat Minkowski background one takes any {\it curved} spacetime
with metric $g_{\mu\nu}$, then the action (\ref{sheet}) changes into
\begin{equation}
\label{curvesheet}
S= -\frac{T}{2} \int d\tau d\sigma \sqrt{-h} h^{ab} g_{\mu\nu}
\partial_{a} X^{\mu} \partial_{b}
X^{\nu} .
\end{equation}
The action (\ref{curvesheet}) is usually called the Polyakov
action \cite{polch}. It is fully equivalent to the so-called
Nambu-Goto action which contains a square root and is simply the
surface area of the string worldsheet
\be
\label{nambugoto}
S =  T \int d\tau d\sigma \sqrt{-h}  .
\ee
It is useful to present the relation between background (target space) metric $g_{\mu\nu}$
and the induced worldsheet metric $h_{ab}$ embedded in $g_{\mu\nu}$
\be
\label{embed}
h_{ab} = g_{\mu\nu} \partial_{a} X^{\mu} \partial_{b} X^{\nu}   .
\ee
In (\ref{curvesheet}) one can then apply the conformal gauge\index{conformal gauge}
\be
\label{confgauge}
\sqrt{-h}h^{ab} = \eta^{ab} ,
\ee
which allows
the 2-dimensional world-sheet metric $h^{ab}$ to be taken as flat metric $\eta^{ab}$.
This is because the action is invariant under Weyl
(conformal) transformations $h^{\prime ab} = f(\sigma)h^{ab}$ and the
$h^{ab}$-dependence can be gauged away. However, Weyl
transformations rescale invariant intervals, hence there is no
invariant notion of distance between two points.
In conformal gauge the action (\ref{curvesheet}) takes the form
\begin{equation}
\label{curvesheetflat}
S= \frac{T}{2} \int d\tau d\sigma \eta^{ab} g_{\mu\nu}
\partial_{a} X^{\mu} \partial_{b}
X^{\nu} .
\end{equation}
In fact, the action (\ref{curvesheetflat}) describes a non-trivial
quantum field theory (QFT), known as nonlinear $\sigma-$model \cite{polch,callan}.

The variation of the action (\ref{curvesheetflat}) gives
equations of motion of a tensile string $(T \neq 0)$ and the
conformal gauge condition (\ref{confgauge}) gives the constraints
equations.

However, the action (\ref{curvesheetflat}) has a disadvantage. Alike the
point particle case with its zero mass limit, one cannot take the
limit of zero tension $T \to 0$ here. In order to avoid this one
has to apply a different action which contains a Lagrange multiplier
$E(\tau,\sigma)$ \cite{alex1,alex2}
\be
\label{alexaction}
S = \frac{1}{2} \int d\tau d\sigma \left[ \frac{g_{\mu\nu}h^{ab} \partial_{a} X^{\mu} \partial_{b}
X^{\nu} }{E^2(\tau,\sigma)} - \frac{E(\tau,\sigma)}{\alpha^{\prime 2}} \right] .
\ee
Varying this action (\ref{alexaction}) with respect to $E$ gives the condition
\be
\label{EOMofE}
E = \alpha' \sqrt{-h}   .
\ee
Substitution (\ref{EOMofE}) back into (\ref{alexaction}) gives
simply the Nambu-Goto action (\ref{nambugoto}).

By the introduction of a new constant $\gamma$ with the dimension
of $(length)^2$ we define a parameter
\be
\varepsilon = \frac{\gamma}{\alpha^{\prime}} .
\ee
Finally, after imposing the gauge
\be
E = - \gamma \left( g_{\mu\nu} X^{\prime \mu} X^{\prime \nu} \right) ,
\ee
together with orthogonality condition
\be
g_{\mu\nu}\dot{X}^{\mu} X^{\prime \nu} = 0 ,
\ee
we get the equations of motion and the constraint for the action (\ref{alexaction})
\cite{dablar,alex,alex1,alex2}
\begin{eqnarray}
\ddot{X}^{\mu}+\Gamma^{\mu}_{\nu\rho}\dot{X}^{\nu}\dot{X}^{\rho} & = &
\varepsilon^2 (X''^{\mu}+\Gamma^{\mu}_{\nu\rho}X'^{\nu}X'^{\rho})\label{ruch1} \:, \\
g_{\mu\nu}\dot{X}^{\mu}\dot{X}^{\nu} & = &
-\varepsilon^2 g_{\mu\nu}X'^{\mu}X'^{\nu} \label{ruch2} ,
\end{eqnarray}
where: $(...)^{.}\equiv\frac{\partial}{\partial\tau}$,
$(...)'\equiv\frac{\partial}{\partial\sigma}$, and $\mu,\nu,\rho=0,1,2,3$
from now on.

Now it makes sense to take the limits:
\begin{itemize}
\item $\varepsilon^2 \to 0$ $(T \to 0)$ for tensionless (null) strings whose worldsheet is
placed on the light cone,
\item $\varepsilon^2 \to 1$ for tensile strings whose worldsheet is placed inside the light cone
\item $\varepsilon = \gamma/{\alpha^{\prime}} \ll 1$ for perturbative scheme for the tensile strings expanded
out of the null strings \cite{alex1,alex2,alex3}.
\end{itemize}
These equations can also be obtained using the gauge as proposed
by Bozhilov \cite{bozhilov}. Another approach to the null string
expansion has been performed in \cite{venico,LS96}.

An important characteristic for both null and tensile strings
is their invariant size defined by (for closed strings) \cite{polch}
\begin{equation}
\label{sizesig}
S(\tau)=\int_0^{2\pi}\;S(\tau,\sigma)\;d\sigma,
\end{equation}
where
\begin{equation}
\label{size}
 S(\tau,\sigma)= \sqrt{-g_{\mu \nu}X^{\prime \mu}X^{\prime\nu}}\; .
\end{equation}

\section{The evolution of strings in black hole spacetimes}
\label{sectionIII}

We start with the study of the evolution of strings in a charged
black hole spacetime or Reissner-Nordstr\"o{}m spacetime which generalizes Schwarzschild
spacetime \cite{chandra}. Reissner-Nordstr\"o{}m spacetime is a spherically symmetric
charged black hole with metric ($t,r,\theta,\phi$ - spacetime coordinates):
\begin{eqnarray}
\label{RNmetric}
ds^{2} & = & (1-\frac{2M}{r}+\frac{Q^{2}}{r^{2}})dt^{2}-(1-\frac{2M}{r}+
\frac{Q^{2}}{r^{2}})^{-1}dr^{2}
 - r^{2}(d\theta^{2}+\sin^{2}\theta d\varphi^{2}) \:,
\end{eqnarray}
where $M$ - mass, $Q$ - charge. In order to get Schwarzschild
black hole one has to put $Q = 0$. For $Q^2 < M^2$ there exist an
event horizon at $r = r_{+} = M + \sqrt{M^2 - Q^2}$ and a Cauchy
horizon at $r = r_{-} = M - \sqrt{M^2 - Q^2}$. For $Q^2 = M^2, r_{+} = r_{-} = M$
and for $Q^2 > M^2$ there are no horizons \cite{he}.

Using the notation: $X^{0}=t(\tau,\sigma)$, $X^{1}=r(\tau,\sigma)$,
$X^{2}=\theta(\tau,\sigma)$, $X^{3}=\varphi(\tau,\sigma)$
the equations of motion for a string in Reissner-Nordstr\"o{}m spacetime are:
\begin{eqnarray}
\label{rneom1} \ddot{t} & - & \varepsilon^{2}
t''+\frac{2M}{r^{3}}(1-\frac{2M}{r}+\frac{Q^{2}}{r^{2}})^{-1}
(r-\frac{Q^{2}}{M})(\dot{r}\dot{t}-\varepsilon^{2} r't')=0 \:,\\
\label{rneom2} \ddot{r} & - & \varepsilon^{2}
r''+\frac{M}{r^{3}}(1-\frac{2M}{r}+\frac{Q^{2}}{r^{2}})
(r-\frac{Q^{2}}{M})(\dot{t}^{2}-\varepsilon^{2} t'^{2})\nonumber\\
& - & \frac{M}{r^{3}}(1-\frac{2M}{r}+\frac{Q^{2}}{r^{2}})^{-1}
(r-\frac{Q^{2}}{M})(\dot{r}^{2}-\varepsilon^{2} r'^{2})\nonumber\\
& - & r(1-\frac{2M}{r}+\frac{Q^{2}}{r^{2}})(\dot{\theta}^{2}-\varepsilon^{2} \theta'^{2})\nonumber\\
& - &
r\sin^{2}\theta(1-\frac{2M}{r}+\frac{Q^{2}}{r^{2}})(\dot{\varphi}^{2}-\varepsilon^{2}\varphi'^{2})=0
 \:,\\
\label{rneom3} \ddot{\theta} & - &
\varepsilon^{2}\theta''+\frac{2}{r}(\dot{r}\dot{\theta}-\varepsilon^{2}
r'\theta')
-\sin\theta\cos\theta(\dot{\varphi}^{2}-\varepsilon^{2} \varphi'^{2})=0 \:,\\
\label{rneom4} \ddot{\varphi} & - &
\varepsilon^{2}\varphi''+\frac{2}{r}(\dot{r}\dot{\varphi}-\varepsilon^{2}
r'\varphi')
+2\cot\theta(\dot{\theta}\dot{\varphi}-\varepsilon^{2}\theta'\varphi')=0\:,
\end{eqnarray}
whereas the constraints are given by
\begin{eqnarray}
&&(1  -  \frac{2M}{r}+\frac{Q^{2}}{r^{2}})\dot{t}^{2}-
(1-\frac{2M}{r}+\frac{Q^{2}}{r^{2}})^{-1}\dot{r}^{2}
-r^{2}(\dot{\theta}^{2}+\sin^{2}\theta\dot{\varphi}^{2})= \nonumber\\
& = & -\varepsilon^{2}\left[(1-\frac{2M}{r}+\frac{Q^{2}}{r^{2}})t'^{2}-
(1-\frac{2M}{r}+\frac{Q^{2}}{r^{2}})^{-1}r'^{2}
-r^{2}(\theta'^{2}+\sin^{2}\theta\varphi'^{2}) \right]\:,\\
& & (1  -  \frac{2M}{r}+\frac{Q^{2}}{r^{2}})\dot{t}t'-
(1-\frac{2M}{r}+\frac{Q^{2}}{r^{2}})^{-1}\dot{r}r'
-r^{2}(\dot{\theta}\theta'+\sin^{2}\theta\dot{\varphi}\varphi')=0\:.
\end{eqnarray}
If one takes $Q = 0$ one gets the equations for the neutral
Schwarzschild spacetime \cite{dablar}.

For a null circular string $(\varepsilon^{2}\rightarrow 0)$ with
circular ansatz:
\begin{eqnarray*}
t=t(\tau), r=r(\tau), \theta=\theta(\tau), \varphi=\sigma,
\end{eqnarray*}
one gets from (\ref{rneom1})-(\ref{rneom4})
\begin{eqnarray}
\label{sc1}
\dot{t} & = & \frac{E(\sigma)}{1-\frac{2M}{r}+\frac{Q^{2}}{r^{2}}}   \:,\\
\label{sc2}
\dot{r}^{2} & - & E^{2}(\sigma)+
\left(1-\frac{2M}{r}+\frac{Q^{2}}{r^{2}}\right)\frac{1}{r^{2}}
\left(K(\sigma)+L^{2}(\sigma)\right)=0\:,\\
\dot{\theta}^{2} & = & r^{-4}\sin^{-2}\theta\left[
K(\sigma)\sin^{2}\theta-L^{2}(\sigma)\cos^{2}\theta\right]\label{nic}\:,\\
\label{sc3}
\dot{\varphi} & = & \frac{L(\sigma)}{r^{2}\sin^{2}\theta}\:,
\end{eqnarray}
where $K(\sigma)$ -- Carter's constant of motion (a constant which refers to coordinate
$\theta$ \cite{dablar}).
It is easy to notice that an energy E, an angular momentum L and
a constant K for a null string do not depend on coordinate $\sigma$.

\subsection{A circular null string with $K=L=0$ in Reissner-Nordstr\"o{}m spacetime}

Firstly, we study the evolution of a null circular string for $K=L=0$. From
(\ref{sc1})-(\ref{sc3}) we obtain
\begin{eqnarray}
\dot{t} & = & \frac{E}{1-\frac{2M}{r}+\frac{Q^{2}}{r^{2}}}\label{osiem} \:,\\
\dot{r}^{2} & = & E^{2} \:,\\
\dot{\theta} & = & 0 \:,\\
\dot{\varphi} & = & 0\label{jedenascie} \:,
\end{eqnarray}
and the constraints are automatically fulfilled.

In analogy to a null circular string that
moves in Schwarzschild spacetime \cite{dablar}, we notice that
Eqs. (\ref{osiem})-(\ref{jedenascie}) describe a "cone" string and its
trajectory is:
\begin{itemize}
\item for $Q^{2}>M^{2}$ given by:
\begin{eqnarray}
& \theta & =const \:,\\
& r & -r_{0}+M\ln|\frac{r^{2}-2Mr+Q^{2}}{r_{0}-2Mr+Q^{2}}|+\nonumber\\
& + &
\frac{2M^{2}-Q^{2}}{\sqrt{Q^{2}-M^{2}}}(arctan\frac{r-M}{\sqrt{Q^{2}-M^{2}}}
-arctan\frac{r_{0}-M}{\sqrt{Q^{2}-M^{2}}})=\nonumber\\
& = & \pm (t-t_{0})\:,
\end{eqnarray}
\item for $Q^{2}<M^{2}$ given by:
\begin{eqnarray}
& \theta & =const \:,\\
& r & -r_{0}+M\ln|\frac{r^{2}-2Mr+Q^{2}}{r_{0}-2Mr+Q^{2}}|+\nonumber\\
& + & \frac{2M^{2}-Q^{2}}{2\sqrt{M^{2}-Q^{2}}}
\ln|\frac{(r-M-\sqrt{M^{2}-Q^{2}})(r_{0}-M+\sqrt{M^{2}-Q^{2}})}
{(r-M+\sqrt{M^{2}-Q^{2}})(r_{0}-M-\sqrt{M^{2}-Q^{2}})}|=\nonumber\\
& = & \pm (t-t_{0})\label{cosik} \:.
\end{eqnarray}
\item for $Q^{2}=M^{2}$ given by:
\begin{eqnarray}
& \theta & =const \:,\\
& r & -r_{0} - M + 2M \ln{\mid \frac{r - M}{r_0 - M} \mid} - \frac{M^2}{r - M}
+ \frac{M^2}{r_0 - M} =  \pm (t-t_{0})\label{cosik1} \:.
\end{eqnarray}
\end{itemize}

\begin{figure}[h]
\centerline{\epsfxsize=10cm\epsfbox{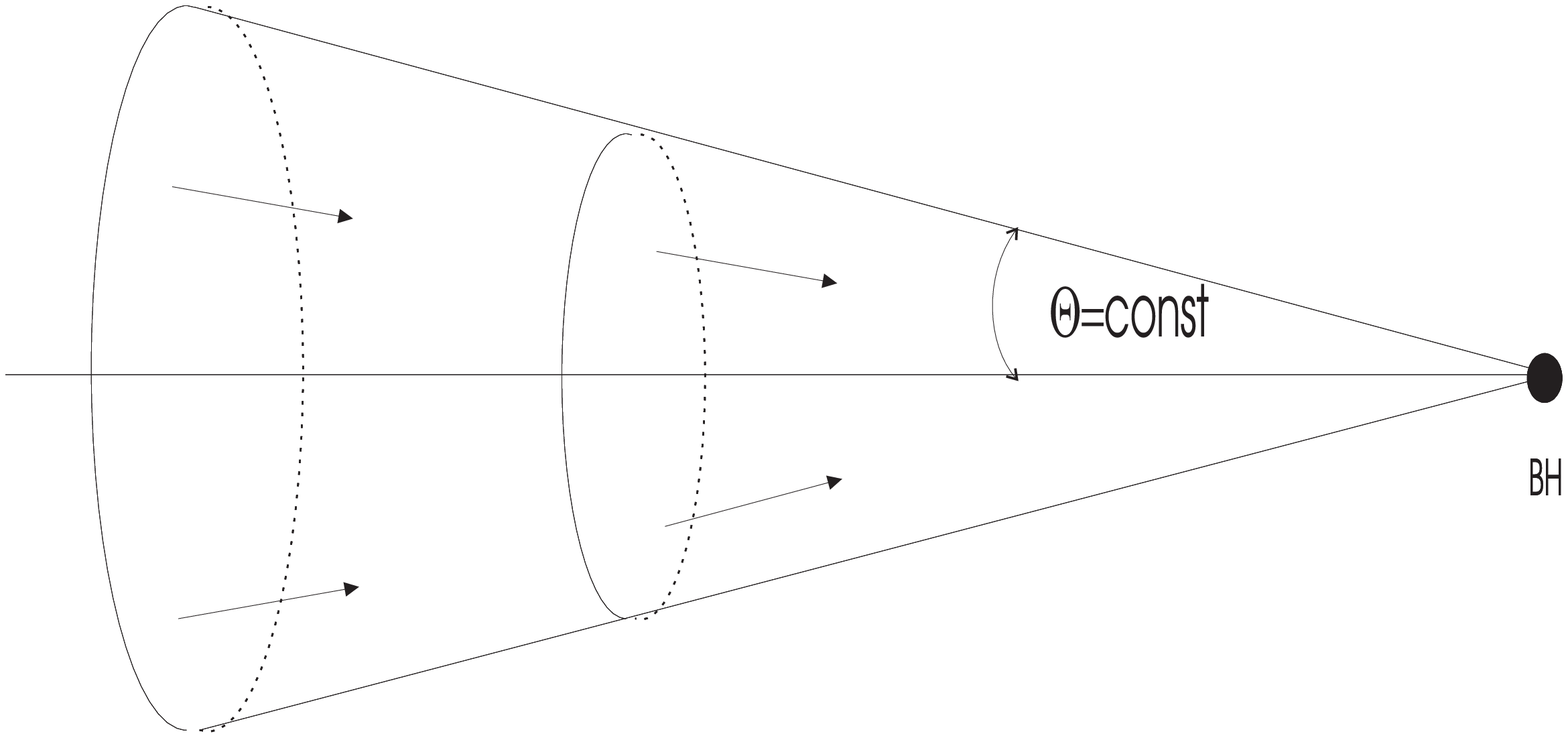}}
\caption{The evolution of a cone string in a black hole (BH) spacetime.}
\label{stozkowa}
\end{figure}

Cone strings start with a finite size
and sweep out a cone of a constant angle $\theta$ (Fig. \ref{stozkowa}).
An observer traveling together with a "cone" string would approach the event horizon
at $r = r_{+}$ after a finite time and then he
would fall onto the singularity (which, in fact, can be escaped of since it is timelike
in Reissner-Nordstr\"om spacetime). On the other hand, an observer at
spatial infinity is not able to notice the moment of passing the
event horizon by the string. The observer sees that the string moves more and more
slowly, in fact, an infinite time to pass the event horizon, or
eventually, fall.

The ''cone'' string is an analogue of a point particle moving on a
radial geodesic, however, it does not move in a
plane through the origin of coordinates $r=0$ but it moves perpendicularly to
the equatorial plane, except for the moment when it is captured. Moreover,
one can find that rotation of such a string is forbidden by the constraints.

Taking the limit $Q=0$ , Eq. (\ref{cosik}) gives exactly the same result
for a cone string as in Schwarzschild spacetime \cite{dablar}. The
equations of motion for Kerr spacetime have been studied in
\cite{porfyriadis98}.

\subsection{A circular null string with $K\neq 0$, $L=0$ in
Reissner-Nordstr\"o{}m spacetime}

Another interesting example of an exact solution is a circular null string
with $K\neq 0$, $L=0$ and the impact parameter $D=3\sqrt{3}M$ (the
impact parameter for strings is defined as $D\equiv
\sqrt{L^{2}+K}/E$) \cite{dablar}. For $D = 3\sqrt{3}M$ there exists
a photon sphere with radius $r_{ph}$ (an unstable photon orbit)
in a Reissner-Nordstr\"o{}m spacetime. In fact, when
\begin{eqnarray}
r &=& r_{ph} = 1,5M\left[1+(1-\frac{8Q^{2}}{9M^{2}})^{\frac{1}{2}}\right]\label{trele}\\
Q^2 &<& \frac{9}{8} M^2 \label{morele}
\end{eqnarray}
one obtains that equations of motion of a string
(\ref{rneom1})-(\ref{rneom4}) are solved by
\begin{eqnarray}
t & = & 3E\tau+\frac{2Q^{2}E\tau}{3M^{2}+3M^{2}
(1-\frac{8Q^{2}}{9M^{2}})^{\frac{1}{2}}-2Q^{2}}\label{jeden} \:,\\
\theta & = & \pm \frac{E\tau}{(1,5M^{2}+1,5M^{2}
(1-\frac{8Q^{2}}{9M^{2}})^{\frac{1}{2}}-Q^{2})^{\frac{1}{2}}}+\theta_{0}\label{dwoje}\:,\\
\varphi &=& \sigma \label{troje} \:.
\end{eqnarray}
The string oscillates an infinite number of times between the
poles of the photon sphere. Its coordinate radius is given by
Eq.(\ref{trele}) with the restriction (\ref{morele}) and its invariant
size (\ref{sizesig}) is given by
\be
S(\tau) = 2
\pi r_{ph} \sin{ \left\{ \pm \frac{E\tau}{ \left[1.5M^2 + 1.5M^2
\left(1 - \frac{8Q^{2}}{9M^{2}} \right)^{\frac{1}{2}} - Q^2
\right]^{\frac{1}{2}}} + \theta_0 \right\}}   .
\ee

In the limit $Q \to 0$ one gets the solution for a string moving on the photon
sphere in schwarzschild spacetime \cite{dablar}. In analogy to a
point particle case one is able to say that these solutions are
unstable with respect to small perturbations.

\begin{figure}[h]
\centerline{\epsfxsize=10cm\epsfbox{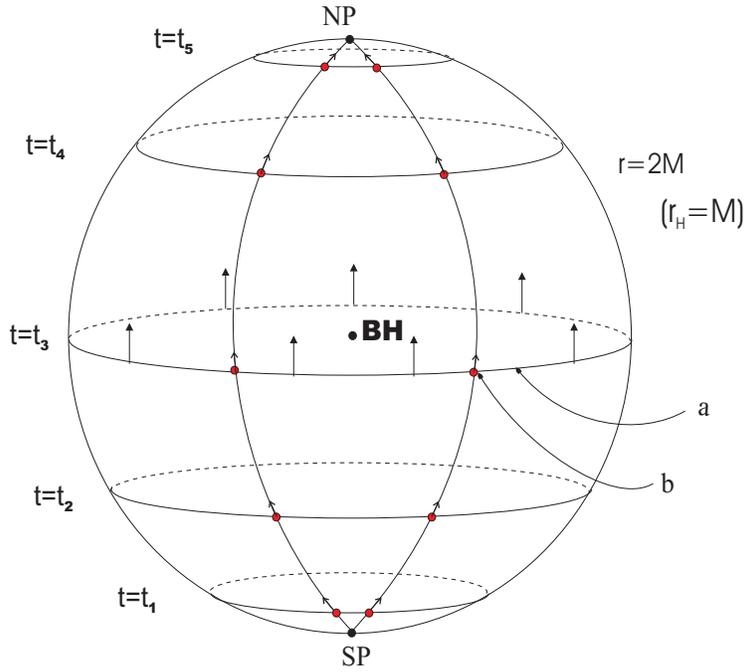}}
\caption{The evolution of a string on the photon sphere in an extreme Reissner-Nordstr\"om
spacetime with $Q^2 = M^2$: a) a string in a moment of passing
the equatorial plane, b) a point of the string moving all the time in
the plane through the origin of coordinate $r=0$. BH -- black hole singularity (here timelike)
, r = 2M -- a radius of the photon sphere, $r_{H} = M$ -- the event horizon.}
\label{sferafot}
\end{figure}

In the special case $Q^2 = M^2$, $r_{ph} = 2M$, $r_+ = M$, the
Eqs.(\ref{jeden})-(\ref{dwoje}) vastly simplify to give (Fig.
\ref{sferafot})
\begin{eqnarray}
\label{Q2M2}
t & = & 4E\tau \:,\\
\theta & = & \pm\frac{E\tau}{M}+\theta_{0} \:,
\end{eqnarray}
and the invariant string size is
\begin{eqnarray}
S(\tau)=4\pi M \sin{\left(\pm\frac{E\tau}{M}+\theta_{0}\right)}\:,
\end{eqnarray}
so that it reduces to zero at poles and to a maximum value at the
equatorial plane. Note that we have to consider the angle $\theta$
as multiply covering angle for Reissner-Nordstr\"om coordinate $\theta$
in metric (\ref{RNmetric}) because the string timelike coordinate
extends from $-\infty < \tau < \infty$.

Let us stress that the solution for a string moving on the photon sphere is not the only
one with a constant $r$. We can find another solution given as
\begin{eqnarray}
t & = & \tau\label{1a} \:,\\
r & = & r_+ = M + \sqrt{M^2 - Q^2} \:,\\
\theta & = & {\rm const.} = \theta_0 \:,\\
\varphi & = & \sigma\label{1b} \:,
\end{eqnarray}
which is analogous to the solution for a null string on the event
horizon \cite{kar} in Schwarzschild spacetime which is a stable solution. Such a string
is placed exactly on the event horizon $r = r_+$. Contrary to a
string moving dynamically on the photon sphere, the string
described by the Eqs.(\ref{1a})-(\ref{1b}) is stationary. Similar
solution exists for a string placed on the Cauchy horizon
\begin{eqnarray}
t & = & \tau\label{2a} \:,\\
r & = & r_- = M - \sqrt{M^2 - Q^2} \:,\\
\theta & = & {\rm const.} = \theta_0 \:,\\
\varphi & = & \sigma\label{2b} \:,
\end{eqnarray}
which is unstable. This is possible since both surfaces of event horizon and Cauchy
horizon are null (isotropic). The problem of evolution of strings
in Reissner-Nordstr\"om spacetime has been studied in both tensile
and null context in Refs. \cite{LS93,LS96,LarS96}. It has been
shown that inside the horizon instabilities appear due to
repulsive effect of a charge. However, for an extreme black hole $(Q^2 = M^2)$
instabilities do not appear.

\section{The evolution of strings in Static Einstein Universe}
\label{sectionIV}

The metric of the static Einstein Universe is \cite{he}:
\begin{eqnarray}
\label{ESUmetric}
ds^{2}&=&dt^{2}-R^{2}\left[\frac{dr^{2}}{1-r^{2}}+r^{2}
\left( d\theta^{2}+\sin^{2}\theta d\varphi^{2}\right)\right]\\
\label{ESUchimetric}
&=& dt^2 - R^2 \left[d\chi^2 + \sin^2{\chi}
\left( d\theta^{2}+\sin^{2}\theta d\varphi^{2}\right)\right]\:,
\end{eqnarray}
where $R=const.$ -- a radius of the universe, $r = \sin{\chi}$ and the proper distance
in the universe is $l = R \chi$, where $0 \le \chi \le \pi$ which corresponds to
$0 \le r \le 1$. The easiest way to study model (\ref{ESUmetric}) is when one introduces
the spherical coordinates
\bea
\label{embedding}
x &=& R\sin{\chi}\sin{\theta}\cos{\phi} \:,\\
y &=& R\sin{\chi}\sin{\theta}\sin{\phi} \:,\\
z &=& R\sin{\chi}\cos{\theta} \:,\\
w &=& R\cos{\chi} \:,
\eea
where $0 \le \theta \le \pi, 0 \le \phi \le 2\pi$. In these
coordinates one is able to embed the 4-sphere $x^2 + y^2 + z^2 + w^2 = R^2$
in a 4-dimensional euclidean space with metric $dS^2 = dx^2 + dy^2 +
dz^2 + dw^2$, or, if one includes a time coordinate, in a
5-dimensional space with metric $d\tilde{S}^2 = -dt^2 + dS^2$.
In such a background the equations of motion for a propagating string are, in
general, given as:
\begin{eqnarray}
\label{ESU1}
\ddot{t} & = & \varepsilon^{2} t'' \:,\\
\label{ESU2}
\ddot{r} & + & \frac{r}{1-r^{2}}\dot{r}^{2}-r(1-r^{2})\dot{\theta}^{2}
-r(1-r^{2})\sin^{2}\theta\dot{\varphi}^{2}=\nonumber\\
& = & \varepsilon^{2}
\left[ r''+\frac{r}{1-r^{2}}r'^{2}-r(1-r^{2})\theta'^{2}
-r(1-r^{2})\sin^{2}\theta\varphi'^{2} \right] \:,\\
\label{ESU3}
\ddot{\theta} & - & \varepsilon^{2}
\theta''+\frac{2}{r}(\dot{r}\dot{\theta}-\varepsilon^{2}r'\theta')
-\sin\theta\cos\theta(\dot{\varphi}^{2}-\varepsilon^{2}\varphi'^{2})=0 \:,\\
\label{ESU4}
\ddot{\varphi} & - &
\varepsilon^{2}\varphi''+\frac{2}{r}(\dot{r}\dot{\varphi}-\varepsilon^{2}
r'\varphi')+2\frac{\cos\theta}{\sin\theta}(\dot{\varphi}\dot{\theta}-\varepsilon^{2}
\varphi'\theta')=0\:.
\end{eqnarray}
The parameter $\varepsilon^{2}$, as before, distinguishes between
null and tensile strings.

\noindent The constraints are
\bea
\label{ESUC1}
\dot{t}^2 &-& \frac{R^2}{1-r^2} \dot{r}^2 - R^2r^2 \dot{\theta}^2 -
R^2 r^2 \sin^2{\theta}\dot{\varphi}^2 = \nonumber \\
&-& \varepsilon^2 \left[
t'^2 - \frac{R^2}{1-r^2} r'^2 - R^2r^2 \theta'^2 -
R^2 r^2 \sin^2{\theta}\varphi'^2 \right] ,\\
\label{ESUC2}
\dot{t}t' &-& \frac{R^2}{1-r^2} \dot{r}r' - R^2r^2 \dot{\theta}\theta' -
R^2 r^2 \sin^2{\theta}\dot{\varphi}\varphi' = 0 .
\eea

\noindent The invariant size (\ref{size}) of a string in the Einstein Static Universe is
given by
\be
S(\tau) = \int_{0}^{2\pi} \left( -t^{\prime 2} +
\frac{R^2}{1-r^2}r'^{2} + R^2 r^2 \theta^{\prime 2} + R^2 r^2
\sin^2{\theta} \varphi^{\prime 2} \right)^{\frac{1}{2}} d\sigma  .
\ee For the null circular string $t = t(\tau), r = r(\tau),
\theta = \theta(\tau), \varphi = \sigma$ one gets \be S(\tau) =
2\pi R r \sin{\theta} .
\ee
First, we consider the following ansatz:
\be
t=t(\tau), r=r(\tau)=\sin{\chi(\tau)}, \theta={\rm const.}, \varphi=\sigma
\ee
(a null circular string with a variable $r$). The solution of the
field equations (\ref{ESU1})-(\ref{ESU4}) is
\begin{eqnarray}
\label{SPI1}
t & = & E\tau \:,\\
\varphi & = & \sigma \:, \\
\theta & = & {\rm const.} = \theta_0 \:,\\
\label{SPI4}
\chi & = & \pm \frac{E\tau}{R}+\chi_{0} \:,
\end{eqnarray}
where we have explicitly used the metric (\ref{ESUchimetric})
and the constraints (\ref{ESUC1})-(\ref{ESUC2}) which are, in fact,
automatically fulfilled. The invariant string size is
\be
S(\tau) = 2\pi R \left[ \sin{\left( \pm \frac{E\tau}{R} + \chi_0
\right)} \right] \sin{\theta_0}.
\ee
The solution (\ref{SPI1})-(\ref{SPI4}) is a cosmological analogue
of the solution (\ref{trele}), (\ref{jeden})-(\ref{troje}) (or in
simpler form (\ref{Q2M2})) which represented a null string on the
photon sphere in Schwarzschild spacetime. It has got the following
physical interpretation: suppose we send a bunch of photons in all
spatial directions from the point $\chi = 0 (r = 0)$ (assuming
that $\chi_0 = 0$) at the moment $t = 0$. These photons form a
spherical plane front of which we consider only a circular bunch
of constant $\theta_0$ - {\it a null string}. The string (the bunch) then starts
from zero size at $\chi = 0 (r=0)$, expands to a maximum size $S=2\pi R\sin{\theta_0}$
what happens for
\be
\tau = \mp \frac{R}{E} \left( \frac{\pi}{2} - \chi_0 \right)
\ee
and finally contracts to zero size again when it reaches the antipodal point at $\chi =
\pi$. Then, the string starts from the antipodal point reaches maximum size and eventually
comes back to an initial point $\chi = 0$ it started with. This
means it returned to the place it was sent after it has traveled
throughout the whole universe. This cycle can then be repeated
infinitely many times. Since the Einstein Static Universe can be
represented as a cylinder in flat space it can be reminded that
each individual point of string will move on a spiral which winds
around this cylinder \cite{he}. Using the embedding equations
(\ref{embedding}) one can show, for instance, that the point $\varphi = \sigma = 0$
is rotating in the hypersurface (x,z,w) while the point $\varphi = \sigma = \pi/2$
is rotating in the hypersurface (y,z,w) etc.

Now, starting with the equations of motion
(\ref{ESU1})-(\ref{ESU4}) we consider a possibility to have
tensile strings ($\varepsilon^2 = 1$) with a constant radial
coordinate $r = \sin{\chi} = const.$ (circular ansatz). Imposing this condition the
equations (\ref{ESU1})-(\ref{ESU4}) simplify to
\bea
\label{constantr1}
\ddot{t} & = & \varepsilon^2 t'' ,\\
\dot{\theta}^2 &+& \sin^2{\theta}\dot{\varphi}^2 = \varepsilon^2
\left[\theta'^2 + \sin^2{\theta} \varphi'^2 \right]  ,\\
\ddot{\theta} &-& \sin{\theta} \cos{\theta} \dot{\varphi}^2 =
\varepsilon^2 \left[\theta'' - \sin{\theta} \cos{\theta} \varphi'^2
\right]  ,\\
\ddot{\varphi} & + & 2\frac{\cos\theta}{\sin\theta}(\dot{\varphi}\dot{\theta}-\varepsilon^{2}
\varphi'\theta')= \varepsilon^2 \varphi''  , \label{constantr4}
\eea
although the constraints (\ref{ESUC1})-(\ref{ESUC2}) do not reduce
so vastly.

The analysis of the equations
(\ref{constantr1})-(\ref{constantr4}) shows that tensile strings
with a constant radial coordinate cannot exist. This is due to
self-interaction of strings (cf. \cite{dablar}).

\section{Strings in Einstein-Schwarzschild spacetime}
\label{sectionV}

In this section we consider the evolution of strings in
Einstein-Schwarzschild (Vadiya) spacetime \cite{vadiya,rumuni}. It describes a point
mass $m$ which is placed in Static Einstein Universe of Section
\ref{sectionIV}. The metric reads as
\be
\label{einschw}
ds^2 = \left(1-\frac{2m}{R}\cot(\frac{r}{R})\right) dt^2 -
\frac{dr^2}{1-\frac{2m}{R}\cot(\frac{r}{R})} - R^2
\sin^2{\left(\frac{r}{R}\right)} \left(d\theta^2 -
\sin^2{\theta}d\varphi^2 \right) .
\ee
It is easy to notice that the coordinate $\chi$ in
(\ref{ESUmetric}) now reads as $\chi = r/R$ and the role of the
radial coordinate similar as in Reissner-Nordstr\"om or Einstein solution is now played by
\be
\bar{r} = \sin{\frac{r}{R}} .
\ee
Other point is that the Einstein metric (\ref{ESUchimetric}) is
obtained in the limit $m \to 0$ while Schwarzschild metric is
allowed in the limit $R \to \infty$. The properties of spacetime
(\ref{einschw}) have been discussed carefully in \cite{rumuni}. It
is interesting to learn that there exist two curvature
singularities: one at $r = 0$ and another at $r = \pi R$. The
former is spacelike in full analogy to Schwarzschild singularity
while the latter is timelike (naked) in analogy to
Reissner-Nordstr\"om singularity of Section \ref{sectionIII}. Therefore, the
metric (\ref{einschw}) describes the Einstein Static Universe with two
antipodal black hole singularities: a spacelike and a timelike.

The equations of motion (\ref{ruch1})-(\ref{ruch2}) for a string moving in the field of metric
(\ref{einschw}) are given by
\begin{eqnarray}
\ddot{t}-\varepsilon^2 t'' & + &
2\frac{m}{R^2\sin^2{(\frac{r}{R})}(1-\frac{2m}{R}\cot(\frac{r}{R}))}(\dot{t}\dot{r}-\varepsilon^2
t'r')=0 \:,\\
\ddot{r}-\varepsilon^2 r'' & + &
\frac{m}{R^2\sin^2{(\frac{r}{R})}(1-\frac{2m}{R}\cot(\frac{r}{R}))}(\dot{r}^2-\varepsilon^2
r'^2)\nonumber\\
& + & \frac{m}{R^2} \frac{\left(1-\frac{2m}{R}\cot(\frac{r}{R})
\right)}{\sin^2{(\frac{r}{R})}} (\dot{t}^2 - \varepsilon^2 t'^2)
\nonumber\\
& - &
R\sin\frac{r}{R}cos\frac{r}{R}(1-\frac{2m}{R}\cot(\frac{r}{R}))(\dot{\theta}^2
-\varepsilon^2\theta'^2) \nonumber\\
 & - & R\sin{(\frac{r}{R})}\cos{(\frac{r}{R})}(1-\frac{2m}{R}\cot(\frac{r}{R}))\sin^2\theta(\dot{\varphi}^2
-\varepsilon^2\varphi'^2)=0 \:,\\
\ddot{\theta}-\varepsilon^2 \theta'' & - &
\sin\theta\cos\theta(\dot{\varphi}^2
-\varepsilon^2\varphi'^2)+\frac{2}{R}\cot(\frac{r}{R})(\dot{r}\dot{\theta}-\varepsilon^2
r'\theta')=0 \:,\\
\ddot{\varphi}-\varepsilon^2 \varphi'' & + &
2\cot\theta(\dot{\varphi}\dot{\theta}-\varepsilon^2
\varphi'\theta')+\frac{2}{R}\cot(\frac{r}{R})(\dot{r}\dot{\varphi}-\varepsilon^2
r'\varphi')=0 \:.
\end{eqnarray}
The constraints read as
\begin{eqnarray}
& - &(1-\frac{2m}{R}\cot(\frac{r}{R}))(\dot{t}^2+\varepsilon^2
t'^2)+(1-\frac{2m}{R}\cot(\frac{r}{R}))^{-1}(\dot{r}^2+\varepsilon^2
r'^2)
+\nonumber\\
& + & R^2\sin^2\frac{r}{R}(\dot{\theta}^2+\varepsilon^2\theta'^2
+\sin^2\theta(\dot{\varphi}^2+\varepsilon^2\varphi'^2))=0\;,\\
& - &(1-\frac{2m}{R}\cot(\frac{r}{R}))\dot{t}t'\nonumber\\
& + &
(1-\frac{2m}{R}\cot(\frac{r}{R}))^{-1}\dot{r}r'+R^2\sin^2\frac{r}{R}(\dot{\theta}\theta'
+\sin^2\theta\dot{\varphi}\varphi')=0 \;.
\end{eqnarray}
For the null strings $(\varepsilon^2 = 0)$, one has
\begin{eqnarray}
\ddot{t} & + &
2\frac{m}{R^2\sin^2{(\frac{r}{R})}(1-\frac{2m}{R}\cot(\frac{r}{R}))}\dot{t}\dot{r}=0 \:,
\label{es1}\\
\ddot{r} & + &
\frac{m}{R^2\sin^2{(\frac{r}{R})}(1-\frac{2m}{R}\cot(\frac{r}{R}))}\dot{r}^2
 + \frac{m}{R^2} \frac{\left(1-\frac{2m}{R}\cot(\frac{r}{R})
\right)}{\sin^2{(\frac{r}{R})}} \dot{t}^2 \nonumber\\
& - &
R\sin{(\frac{r}{R})}\cos{(\frac{r}{R})}\left(1-\frac{2m}{R}\cot(\frac{r}{R})\right)\left(\dot{\theta}^2
+\sin^2\theta\dot{\varphi}^2\right)=0 \:, \label{es2}\\
\ddot{\theta} & - & \sin{\theta}\cos{\theta}\dot{\varphi}^2
+\frac{2}{R}\cot(\frac{r}{R})\dot{r}\dot{\theta}=0 \:, \label{es3}\\
\ddot{\varphi} & + &
2\cot\theta\dot{\varphi}\dot{\theta}+\frac{2}{R}\cot(\frac{r}{R})\dot{r}\dot{\varphi}=0
\:,\label{es4}
\end{eqnarray}
and
\begin{eqnarray}
& -&(1-\frac{2m}{R}\cot(\frac{r}{R}))\dot{t}^2+(1-\frac{2m}{R}\cot(\frac{r}{R}))^{-1}\dot{r}^2
+\nonumber \\
& + & R^2\sin^2\frac{r}{R}(\dot{\theta}^2
+\sin^2\theta\dot{\varphi}^2)=0\;, \label{wes1}\\
& - &(1-\frac{2m}{R}\cot(\frac{r}{R}))\dot{t}t'\nonumber\\
& + &
(1-\frac{2m}{R}\cot(\frac{r}{R}))^{-1}\dot{r}r'+R^2\sin^2\frac{r}{R}(\dot{\theta}\theta'
+\sin^2\theta\dot{\varphi}\varphi')=0 \;. \label{wes2}
\end{eqnarray}
The first integrals of (\ref{es1})-(\ref{es4}) are (compare
\cite{dablar})
\bea
\dot{t} &=& \frac{E(\sigma)}{1-\frac{2m}{R}\cot(\frac{r}{R})} \;, \label{esint1}\\
\dot{\varphi}&=&\frac{L(\sigma)}{sin^2\theta\sin^2{(\frac{r}{R})}}
\;, \label{esint2}\\
\sin^4{(\frac{r}{R})} \sin^2{\theta} \dot{\theta}^2 &=& -
L^2(\sigma) \cos^2{\theta} + K(\sigma) \sin^2{\theta} \;,
\label{esint3}
\eea
and
\be
\dot{r}^2 + V(r) = 0 \;,
\ee
where
\be
V(r) = -E^2(\sigma) + \frac{R^2}{\sin^2{(\frac{r}{R})}}
\left(1 -\frac{2m}{R}\cot{(\frac{r}{R})} \right) \left( L^2(\sigma) +
K(\sigma) \right) \;.
\ee

There exists a solution with a constant $r$ given by
\begin{eqnarray}
t & = & \tau\label{10a} \:,\\
r & = & r_H = R\hspace{0.1pt} \arctan{\left( \frac{2m}{R}\right)} \:,\\
\theta & = & {\rm const.} = \theta_0 \:,\\
\varphi & = & \sigma\label{10b} \:,
\end{eqnarray}
which is analogous to the solution for a null string on the event
horizon (so that it should be stable) in the Reissner-Nordstr\"om spacetime given by Eqs. (\ref{1a})-(\ref{1b}).
Here the event horizon is at $r_H$.

It is interesting to notice that apparently there should exist another solution
with a constant $r$ which would be for the photon sphere $r_{ph}$ in Einstein-Schwarzschild
spacetime given by
\begin{eqnarray}
t & = & 3E\tau\label{11a} \:,\\
r & = & r_{ph} = R\hspace{0.1pt} \arctan{\left( \frac{3m}{R}\right)} \:,\\
\theta & = & \frac{E\tau}{\sqrt{3} m} \sqrt{1 + \frac{9m^2}{R^2}}:,\\
\varphi & = & \sigma\label{11b} \:,
\end{eqnarray}
which would be analogous to the solution for a null string on the photon
sphere in Schwarzschild spacetime \cite{dablar} which can be obtained in the limit
$R \to \infty$ (or by taking $Q \to 0$ in the solution (\ref{1a})-(\ref{1b})
for a null string on the photon sphere in Reissner-Nordstr\"om spacetime).
However, it is a simple exercise to show that {\it this solution is a
contradiction}, namely there is a conflict between the field
equation (\ref{es1}) and the constraint (\ref{wes1}). The physical
reason for this is similar to those which produces stationary
strings in the de Sitter spacetime (cf. \cite{multi}) - since
there is no string tension which can balance local gravity a
stationary or better static string cannot exist.

On the other hand, there exists a solution for a ``cone'' string given by
\bea
t & = & E \int{\frac{d \tau}{1 - \frac{2m}{R} \cot{(\pm \frac{E}{R} \tau)}}} \:,\\
r & = & \pm E \tau \:,\\
\theta & = & {\rm const.} =  \theta_0 \:,\\
\phi & = & \sigma \:.
\eea
This is analogous to the solution (\ref{SPI1})-(\ref{SPI4}) in
Einstein Static Universe. It is also easy to prove in the
same way as in Schwarzschild/Reissner-Nordstr\"om and Einstein
spacetimes that there exist no tensile circular strings of constant
radius.

\section{Conclusion}
\label{concl}

In this paper we have found some exact string configurations in black hole and cosmological
spacetimes which apply both for fundamental and for cosmic strings.
We generalized previously found solutions of Ref. \cite{dablar} for a "cone" string
and for a string moving on the photon sphere into a Reissner-Nordstr\"om
spacetime which is also related to the discussion of the behaviour
of strings in this spacetime given in refs. \cite{LS93,LS96,LarS96}.
We also generalized an event horizon solution and presented a
Cauchy horizon solution for the Reissner-Nordstr\"om spacetime.
We found a solution for a null string moving around the Einstein Static Universe
and two completely new solutions for strings evolving in the
Einstein-Schwarzschild spacetime (a black hole in the Einstein
Static Universe).

Firstly, we briefly presented formalism which allowed to take the limit of null
strings in an appropriate action. Then, we studied the evolution of
strings in Reissner-Nordstr\"om, Einstein Static and
Einstein-Schwarzschild spacetimes. The exact configurations we
found can be grouped geometrically into a couple of classes. There is a class of
solutions which describe null strings residing on the null
surfaces of these spacetimes, i.e., event and Cauchy horizons.
There is also a class of solutions which describe strings
sweeping out the light cones of a particular spacetime. Another class
is for strings which reside on the surface of the photon sphere
(an unstable periodic orbit for zero point particles). This class
exists both in Schwarzschild/Reissner-Nordstr\"om spacetimes and,
in an adapted form, in the Einstein Static Universe, but not in the
Einstein-Schwarzschild spacetime.

As far as the physical properties are concerned we found that some
of our solutions are unstable (for instance, a string on the photon
sphere in Reissner-Nordstr\"om spacetime) and some are stable
(e.g., a string on the event horizon). According to Ref.
\cite{multi}, multistring solutions appear whenever the
world-sheet time $\tau$ is a multi-valued function of the physical
time and they are possible, for instance, in the positive
cosmological constant models such as the de Sitter space. In our
paper only the Einstein Static Umiverse admits a positive
cosmological constant and because of that one should perhaps
expect some multistring solutions admissible. However, our
solutions of Sections IV and V do not possess this property. On
the other hand, some of our solutions (a string on the photon
sphere (Eqs. (\ref{jeden})-\ref{troje})) and a string in the Einstein Static Universe
(Eqs. (\ref{SPI1})-(\ref{SPI4})) have an
invariant string size described by multiply covering azimuthal
angle because of an infinite domain of the timelike string
coordinate $\tau$.

The existence of the photon sphere, i.e., an unstable periodic orbit
(UPO), together with other special solutions suggests that a general
evolution of a tensile (or perhaps even a null) string in these
simple curved backgrounds is chaotic.
This statement is obviously true for Schwarzschild spacetime
\cite{frolar}, and the solutions we have found are straightforward generalizations
of exact configurations in Schwarzschild spacetime.

The results we gained can give some insight into the nature of
motion of strings in extremely high gravitational fields of black
holes and in the early universe in fully quantum string theory.

\section{Acknowledgements}

We thank Franco Ferrari, Alexander Kapustnikov, Arne Larsen, Sergey Roshchupkin and
Alexander Zheltukhin for useful discussions. MPD was
partially supported by the Polish Research Committee (KBN) Grant No 2PO3B
105 16.

\end{document}